\pdfoutput=1

\documentclass[11pt]{article}

\usepackage[]{ACL2023}

\usepackage{times}
\usepackage{latexsym}
\usepackage{subcaption}

\usepackage[T1]{fontenc}

\usepackage[utf8]{inputenc}

\usepackage{microtype}

\usepackage{inconsolata}

\usepackage{hyperref}
\usepackage{multirow}
\usepackage{graphicx}
\usepackage{tabularx}
\usepackage{url}
\usepackage{xcolor}
\usepackage{epsfig}
\usepackage{adjustbox}
\usepackage{amsfonts, amsmath, amssymb}
\usepackage{booktabs} 
\usepackage{comment}
\usepackage{caption}
\usepackage{subcaption}
\usepackage{textcomp}
\usepackage{relsize}
\usepackage{stmaryrd}
\usepackage{bbm}
\usepackage{rotating}
\usepackage{helvet}
\usepackage{courier}
\usepackage{cleveref}
\usepackage{xspace}
\usepackage{enumitem}
\usepackage{epigraph}
\usepackage{listings}
\usepackage{mdframed}
\usepackage{makecell}
\usepackage{tcolorbox}
\usepackage{nicematrix}
\usepackage{tabu}
\usepackage{colortbl}
\usepackage{xcolor}
\usepackage{array}
\PassOptionsToPackage{table}{xcolor}

\usepackage{algorithm}
\usepackage{algpseudocode}

\usepackage{amsthm}

\definecolor{codegreen}{rgb}{0,0.6,0}
\definecolor{codegray}{rgb}{0.5,0.5,0.5}
\definecolor{codepurple}{rgb}{0.58,0,0.82}
\definecolor{backcolour}{rgb}{0.95,0.95,0.92}

\lstdefinestyle{pythonstyle}{ 
    commentstyle=\color{codegreen},
    keywordstyle=\color{magenta},
    stringstyle=\color{codepurple},
    basicstyle=\ttfamily\small,
    breakatwhitespace=false,         
    breaklines=true,                 
    captionpos=b,                    
    keepspaces=true,
    numbersep=5pt,                  
    showspaces=false,                
    showstringspaces=false,
    showtabs=false,                  
    tabsize=2,
    language=Python
}

\theoremstyle{definition}
\newtheorem{definition}{Definition}[section]

\newcolumntype{L}{>{\arraybackslash}m{0.99\linewidth}}

\tcbset{
  aibox/.style={
    width=\textwidth,
    top=10pt,
    colback=white,
    colframe=black,
    colbacktitle=black,
    center title,
    fonttitle=\bfseries,
  }
}

\newtcolorbox{AIbox}[2][]{aibox,title=#2,#1}

\tcbset{
  aiboxsmall/.style={
    width=0.98\linewidth,
    top=10pt,
    colback=white,
    colframe=black,
    colbacktitle=black,
    center title,
    fonttitle=\bfseries,
  }
}

\newtcolorbox{AIboxSmall}[2][]{aiboxsmall,title=#2,#1}

\title{How Diversely Can Language Models Solve Problems? Exploring~the~Algorithmic Diversity of Model-Generated Code}

\author{Seonghyeon Lee\textsuperscript{1}, Heejae Chon\textsuperscript{3}, Joonwon Jang\textsuperscript{2}, Dongha Lee\textsuperscript{3}, Hwanjo Yu\textsuperscript{2} \\
  \textsuperscript{1}School of Computer Science and Engineering, Kyungpook National University, South Korea\\
  \textsuperscript{2}Department of Computer Science and Engineering, POSTECH, South Korea\\
  \textsuperscript{3}Department of Artificial Intelligence, Yonsei University, South Korea\\
  \texttt{sh0416@knu.ac.kr}}

\begin{document}
\maketitle
\begin{abstract}
Language models (LMs) have exhibited impressive abilities in generating code from natural language requirements. 
In this work, we highlight the diversity of code generated by LMs as a critical criterion for evaluating their code generation capabilities.
There is a lack of studies focused on assessing the diversity of generated code, which overlooks its importance in code LMs.
Therefore, we propose a systematic approach to evaluate code diversity, introducing various metrics with inter-code similarity. 
Specifically, we introduce code clustering methods that leverages LMs' capabilities in code understanding and reasoning, resulting in a set of metrics that represent the number of algorithms in model-generated solutions.
We extensively investigate the property of model-generated solutions by contrasting them with human-written ones and quantifying the impact of various factors on code diversity: model size, temperature, instruction tuning, and problem complexity.
Our analysis demonstrates that model-generated solutions exhibit low algorithmic diversity, which was neglected by the research community.
Moreover, we explore methods to increase code diversity by combining solutions from different models and increasing sampling temperatures.
Our findings highlight that code diversity can be enhanced with the help of heterogeneous models and setting temperature beyond 1.0 that has not been fully explored due to the functional correctness degradation.
To facilitate our research direction, we publicly share our code and datasets through open-source repositories.





\end{abstract}

\section{Introduction}
\label{sec:intro}
\epigraph{There's more than one way to do it.}{\textit{A motto for the Perl programming language \\ Larry Wall}}

Language models (LMs) should produce \textit{correct} and \textit{diverse} implementations~\citep{li-etal-2016-diversity,le-bronnec-etal-2024-exploring,tekin-etal-2024-llm}.
Recently, language models that generate code, called Code LMs, have garnered attention due to their practical availability.
Code LMs~\citep{guo2024deepseekcoder,codegemmateam2024codegemmaopencodemodels,rozière2024codellamaopenfoundation,lozhkov2024starcoder} mainly focus on enhancing their correctness with a single solution as the current state of code generation evaluation systems only measures the functional correctness~\citep{austin2021program, chen2021evaluating, wang2022mconala,bigcode-evaluation-harness}.
However, the diversity of generated code is another important criterion for measuring their potentials.
For instance, models tend to produce correct code when they are allowed to generate a large number of candidate implementations.
Therefore, researchers promote \textit{diversity} by adjusting sampling temperature to increase the probability of generating \textit{correct} code~\citep{renze-2024-effect}.
It indicates that diversity is one crucial aspect to identify the hidden value in code LMs.
In addition, the diversity of generated code represents their ability to come up with creative ideas for a problem and instantiate them through programming languages, which reflects the core intellectual property for programming.

Nevertheless, there exist limited studies focusing on analyzing the diversity of generated code.
In the literature on evaluating the capabilities of code LMs,  \citet{hendrycks2021measuring,yadav-etal-2024-pythonsaga,matton-etal-2024-leakage} focus only on functional correctness without exploring various implementation patterns in the generated code.
Recently, \citet{zhuo-2024-ice,tong-zhang-2024-codejudge} introduce automatic metrics to evaluate several other aspects such as usefulness of generated code, but they do not measure how much the model-generated solutions vary algorithmically.
This tendency overlooks the potential ability of code LMs and hinders the opportunity for LMs to learn to think of diverse ideas when generating code.

In this work, we aim to systematically investigate algorithmic diversity encoded in code LMs with the following research questions:
\begin{itemize}[leftmargin=*,topsep=2pt,itemsep=2pt,parsep=0pt]
\item \textbf{Q1}: How can we design a quantifiable metric for \textit{algorithmic diversity} of model-generated code?
\item \textbf{Q2}: How different is \textit{algorithmic diversity} between human-written and model-generated code? 
\item \textbf{Q3}: Which factors in model development, e.g., instruction tuning or quantization, affect \textit{algorithmic diversity} of generated solutions?
\end{itemize}

To this end, we propose clustering-based evaluation metrics to estimate the number of algorithms within a set of solutions.
Our approach leverages Large Language Models (LLMs) to determine the algorithmic equivalence between two different implementations through reasoning.
With these predictions, solutions are clustered based on their algorithmic similarity, resulting in two distinct metrics that quantify the number of unique algorithms: $\mathrm{DA@}K$ and $\mathrm{EA}$.
When further considering the relationship between the number of solutions and the number of algorithms, we introduce an Algorithmic Diversity Curve along with its corresponding metric, i.e., $\mathrm{NAUADC}$.

Our experimental results provide new insights in Code LMs that were previously unavailable.
First, we found that human-generated solutions are more diverse than the correct solutions generated by language models.
Our algorithmic diversity curve analysis shows that although models generate many correct solutions, they contain fewer unique algorithms than humans.
In addition, we explore the impact of different model series, model size, sampling temperature, and instruction tuning in terms of diversity.
Our experimental results show a different algorithmic diversity trend even though they show similar functional correctness scores.
Lastly, we explore several approaches to enhance the diversity in a solution pool by using merged solution and high temperature sampling, suggesting the importance of introducing heterogeneous models and exploring their generation space more aggressively through high-temperature sampling.

\section{Related work}
\label{sec:relwork}
\subsection{Code Language Model Evaluation}
Evaluating language models' code generation capabilities has become an intriguing research topic in our community~\citep{chen2021evaluating,bigcode-evaluation-harness}.
Since code solutions can be readily validated through test case execution, recent studies~\citep{NEURIPS2023_43e9d647,wang-etal-2023-recode} have primarily focused on functional correctness, overlooking other important aspects of generated code.
There exist few studies to fuse fuzzy metrics with the syntactic code structure such as AST~\citep{ren2020codebleumethodautomaticevaluation} or capture the code semantics with masked language models, i.e., CodeBERTScore~\citep{,zhou-etal-2023-codebertscore}, but they often failed to effectively represent code semantics.
Recently, as incorporating LMs into general evaluation pipeline boosts the evaluation quality~\citep{liu-etal-2023-g,fu-etal-2024-gptscore}, \citet{zhuo-2024-ice,tong-zhang-2024-codejudge} introduce LMs to evaluate the code semantics, e.g., ICE Score or CodeJudge.

Along with this literature, our work first presents an effective methodology to analyze the code diversity through LMs in terms of algorithmic behavior.
This aspect has not been fully explored by the community, and we comprehensively investigate code diversity embedded within code LMs.

\subsection{Diversity Evaluation}
Diversity acts as a core virtue for building language generation systems~\citep{li-etal-2016-diversity}.
It is a core factor for constructing general instruction-tuned language models~\citep{xu2024wizardlm,bukharin-etal-2024-data,shypula2025does}.
These trends have also been verified for code LMs~\citep{pmlr-v235-wei24h,luo2024wizardcoder,wang-etal-2024-dolphcoder}.
However, there exists limited progress in evaluating the diversity encoded in code LMs.

Properly evaluating the diversity has been a long-standing research topic~\citep{tevet-berant-2021-evaluating,NEURIPS2021_260c2432,lu2024benchmarkinglanguagemodelcreativity}.
To do this, we often utilize the predicted relevance between two texts such as sentences or code snippets~\citep{gao-etal-2021-simcse,lee-etal-2022-toward,feng-etal-2020-codebert,lu2021codexglue,zakeri2023systematic}.
Particularly, several approaches have been developed to capture code relevance, e.g., token-based~\citep{7886988}, learning-based~\citep{10.1016/j.asoc.2022.109562}, or hybrid techniques~\citep{8816761}.
However, they failed to properly capture the semantic relevance in coding domain due to their complexity.

Building on this line of research, we introduce LMs' reasoning ability for measuring the diversity of code LMs.
Moreover, we integrate a clustering method with this similarity to derive a intuitive measures for representing code diversity.



\section{Algorithmic Diversity Metrics}
\label{sec:eval}
In this section, we introduce clustering-based metrics that measure algorithmic diversity of solutions.

\subsection{Preliminary}
\paragraph{Inferring algorithmic discrepancy between two codes through LLMs}
We design a reasoning-based similarity measure, named $\mathrm{AlgoSim}$, to effectively capture the algorithmic discrepancy between two codes.
Motivated by \citet{fu-etal-2024-gptscore}, we utilize the instruction-tuned language models to explain their algorithmic behaviors separately and then determine their algorithmic difference conditioned on self-generated explanations.
The models are prompted to produce a binary decision, i.e., similar or different.
Our prompt is presented in Appendix~\ref{app:prompt}.

\paragraph{Clustering solutions based on algorithmic difference}
We introduce a simple yet effective clustering algorithm to express the diversity inside a set of solutions.
Specifically, given a sufficient number of codes for a problem, we initialize a set of solution clusters and iteratively update this set by comparing each solution with each cluster.
If the current solution introduces a new algorithm that is different from those of existing clusters, then we create a new cluster for this solution in the set.
Otherwise, we assign the solution into the cluster that has a similar algorithm.
We outline the overall procedure in Algorithm~\ref{alg:cluster}.

\begin{algorithm}[t]
\caption{Solution clustering algorithm}\label{alg:cluster}
\begin{algorithmic}
\Require A set of solutions $\mathcal{S} = \{s_1,...s_N\}$
\Ensure $\mathcal{C} = \{ \mathcal{C}_1, ..., \mathcal{C}_M \}$ s.t. $\mathcal{C}_i = \{ s_1,...,s_j\}$
\State $\mathcal{C} \gets \emptyset$
\While{$\mathcal{S} \neq \emptyset$}
\State $s \sim \mathcal{S}$ \Comment{Sample a solution}
\State $\mathcal{S} \gets \mathcal{S} \setminus \{ s \}$, $\mathcal{C}' \gets \{ s \}$ \Comment{Initialize a cluster}
\ForAll{$s' \in \mathcal{S}$}
\If{$\mathrm{AlgoSim}(s,s') = \mathrm{Same}$}
\State $\mathcal{S} \gets \mathcal{S} \setminus \{ s' \}$, $\mathcal{C}' \gets \mathcal{C}' \cup \{ s' \}$
\EndIf
\EndFor
\State $\mathcal{C} \gets \mathcal{C} \cup \{ \mathcal{C}' \}$  \Comment{Insert new cluster}
\EndWhile
\end{algorithmic}
\end{algorithm}

\subsection{M1: The Number of Distinct Algorithms}
We quantify algorithmic diversity by estimating \textbf{the number of distinct algorithms (DA)} within a given solution set of size $N$.  
Conceptually, DA corresponds to the number of solution clusters in the clustering results.  
Thus, we define $\mathrm{DA}$ as the total count of these clusters for a problem.  
However, in practice, the number of available solutions ($N$) varies across different problems, affecting the estimation of $\mathrm{DA}$.
For instance, introductory problems often have a large pool of solutions, while more challenging problems tend to have fewer.

\paragraph{Decoupling the available solution set size $N$ and the evaluation set size $K$}
We derive an efficient estimate for $\mathrm{DA@}K$ given an available solution set with size $N$, where $K$ and $N$ are typically different.
A straightforward approach is to use Monte Carlo estimation by sampling $K$ solutions from the set.
However, achieving accurate estimates requires multiple sampling iterations, making it computationally expensive.
To address this, we propose an efficient and accurate method to estimate $\mathrm{DA}@K$ directly from a solution set of size $N$, eliminating the need for repeated sampling.
\begin{definition}
\label{def:da@k_estimation}
An efficient estimate of DA@$K$ from a solution set of size $N$ is as follow:
\begin{equation}
\mathrm{DA@}K = \sum_{m=1}^M \left(1 - \frac{{N-s_m \choose K}}{{N \choose K}} \right)
\end{equation}.
where $M$ is the number of solution clusters and $s_m$ is the size of $m$-th cluster.
\end{definition}

To derive this estimate, we interpret $\mathrm{DA@}K$ as the expected count of clusters that contribute at least one solution within the sampled set of size $K$.  
By leveraging the linearity of expectation, this simplifies to the sum of the probabilities that each cluster is represented in the sampled subset.
\begin{equation*}
\mathrm{DA@}K = \sum_{m=1}^M \mathbb{E}\left[ p \left( ^\exists s | s \in \mathcal{S}_K \land s \in \mathcal{C}_m \right) \right]
\end{equation*}
where $\mathcal{S}_K \sim \mathcal{S}$ is a sampled solution set with size $K$ and $\mathcal{C}_m$ is the $m$-th cluster.
To compute the expectation for group $m$, we follow these steps:
\begin{enumerate}[leftmargin=*,topsep=2pt,itemsep=2pt,parsep=0pt]
\item We calculate the probability of not selecting the group. It can be calculated as the ratio between two values: the number of sampled sets that do not contain any solution in the group $m$ and the total number of possible sampled sets.
\item We take the complement of this probability to obtain the likelihood of sampling at least one solution in each group.
\end{enumerate}
Finally, we add each expectation to obtain $\mathrm{DA@}K$.

\subsubsection{Algorithmic Diversity Curve}
We develop the Algorithmic Diversity Curve (ADC) to express the trend between the number of solutions and our proposed metrics.
The curve plots the relationship between the number of solutions ($K$) on the horizontal axis and $\mathrm{DA@}K$ values on the vertical axis.
To create the ADC, we generate a grid of $K$ values and estimate the $\mathrm{DA@}K$ at each point using Definition~\ref{def:da@k_estimation}.
Motivated by MAUVE~\citep{NEURIPS2021_260c2432}, we introduce Normalized Area Under ADC (NAUADC) as a generalized metric that captures the average number of distinct algorithms across different solution set sizes.
To calculate NAUADC, we measure the area under ADC and normalize it with respect to the width of the curve.

\subsection{M2: The Effective Number of Algorithms}
We use the concept of uncertainty to measure algorithmic diversity.
Motivated by the concept of diversity index~\citep{spellerberg2003tribute} and semantic uncertainty~\citep{kuhn2023semantic}, we define \textbf{the effective number of algorithms~(EA)} using the following equation:
\begin{equation}
\mathrm{EA} = \exp \left( - \sum_{m=1}^M p_m \ln p_m \right)
\end{equation}
where $p_m:=\frac{|\mathcal{C}_m|}{N}$ is the empirical estimate of implementing a solution with algorithm $m$.
Unlike $\mathrm{DA}@K$ which counts unique algorithms, $\mathrm{EA}$ examines the distribution pattern of diverse implementations across the solution set.
This helps us understand how evenly the model implements various algorithms.
For instances, $\mathrm{EA}$ will be low if solutions predominantly use the same algorithm, even if the model can occasionally generate novel approaches across multiple attempts.

\section{Experiments}
\label{sec:exp}
We analyze the algorithmic diversity encoded in model-generated solutions and discover their potential in code generation task.

\subsection{Experimental Setup}

\paragraph{General settings}
We investigate algorithmic diversity of generated solutions using the APPS dataset~\citep{hendrycks2021measuring}.
We group the problems by their original sources as the number of collected human-written solutions varies (Table~\ref{tab:exp1-data-stats}).
We select two sources, i.e., AtCoder and CodeForces, which contain sufficient number of problems and human solutions per difficulty bin.
For model-generated solutions, we employ state-of-the-art code LMs including Deepseek Coder~\citep{guo2024deepseekcoder}, Qwen2.5 Coder~\citep{hui2024qwen25codertechnicalreport} with their quantized versions~\citep{lin2023awq} and proprietary models, i.e., OpenAI GPT models~\citep{openai2024gpt4ocard}.
With these models, we use sampling temperature 1.0 with nucleus sampling 0.95 and generate at most 1024 tokens until EOS token is produced.
The open-sourced models generate 100 solutions per problem and the proprietary models generate 20 solutions per problem due to their expensive costs.
We extract code solutions from the completions using a simple regex pattern that captures the first markdown codeblock.
The generated solutions are executed with test cases to verify their functional correctness.

\paragraph{Settings for diversity metrics}
For inferring $\mathrm{AlgoSim}$, we employ the Llama-3.1-8B-Instruct model~\citep{grattafiori2024llama3herdmodels} to assess algorithmic relevance between two implementations. 
This model is chosen for its strong instruction-following capabilities while remaining computationally feasible on a single NVIDIA A100 GPU.
We vary $K$ from 1 to 25 to draw the ADC curve and analyze the trends of $\mathrm{DA@}K$ as the solution set size increases. 
The $\mathrm{NAUADC}$ is calculated based on these settings.

\begin{table}[]
\small
\centering
\begin{tabular}{@{}llcc@{}}
\toprule
\multicolumn{1}{c}{Difficulty} & \multicolumn{1}{c}{Source} & Problem & Solution \\ \midrule
\multirow{2}{*}{\textit{Introductory}} & AtCoder & 403 & 97.57±6.56 \\
 & CodeForces & 294 & 18.07±7.69 \\ \midrule
\multirow{2}{*}{\textit{Interview}} & AtCoder & 252 & 82.46±25.36 \\
 & CodeForces & 2376 & 17.73±9.00 \\ \midrule
\multirow{2}{*}{\textit{Competition}} & AtCoder & 41 & 34.51±35.80 \\
 & CodeForces & 264 & 14.26±9.45 \\ \bottomrule
\end{tabular}
\caption{Statistics of the APPS test dataset. We report the number of problems and the average (w/ stdev) number of human-written solutions for a single problem.}
\label{tab:exp1-data-stats}
\end{table}

\begin{table*}[]
\centering
\resizebox{\textwidth}{!}{%
\begin{tabular}{@{}lccccccccc@{}}
\toprule
\multicolumn{1}{c}{Difficulty} & \multicolumn{3}{c}{\textit{introductory}} & \multicolumn{3}{c}{\textit{interview}} & \multicolumn{3}{c}{\textit{competition}} \\ \midrule
\multicolumn{1}{c}{Model} & Pass@10 & EA & NAUADC & Pass@10 & EA & NAUADC & Pass@10 & EA & NAUADC \\ \midrule
\multicolumn{10}{c}{Source: AtCoder} \\ \midrule
Deepseek-6.7B-Base & 0.6683 & \textbf{1.653} & \textbf{1.900} & 0.1313 & 1.354 & 1.454 & 0.0073 & 1.000 & 1.000 \\
Deepseek-6.7B-Instruct & 0.7637 & 1.536 & 1.747 & 0.2114 & 1.578 & 1.733 & 0.0155 & \textbf{1.333} & \textbf{1.321} \\
Deepseek-6.7B-Instruct-AWQ & 0.7361 & 1.516 & 1.741 & 0.1798 & 1.484 & 1.637 & 0.0046 & 1.000 & 1.000 \\
Deepseek-33B-Instruct-AWQ & 0.8772 & 1.548 & 1.780 & 0.3223 & \textbf{1.607} & \textbf{1.814} & 0.0097 & 1.000 & 1.000 \\
GPT-4o-mini-2024-07-18 & 0.9219 & 1.283 & 1.443 & 0.4518 & 1.293 & 1.444 & \textbf{0.1279} & 1.211 & 1.350 \\
GPT-4o-2024-08-06 & \textbf{0.9367} & 1.216 & 1.302 & \textbf{0.5612} & 1.333 & 1.409 & 0.1185 & 1.137 & 1.136 \\ \midrule
Human &  & 2.080 & 2.353 &  & 2.817 & 3.042 &  & 2.484 & 2.722 \\ \midrule
\multicolumn{10}{c}{Source: CodeForces} \\ \midrule
Deepseek-6.7B-Base & 0.1100 & 1.382 & 1.496 & 0.1234 & 1.458 & 1.573 & 0.0495 & 1.418 & 1.461 \\
Deepseek-6.7B-Instruct & 0.1993 & 1.462 & 1.643 & 0.2215 & 1.548 & 1.714 & 0.0991 & 1.438 & 1.555 \\
Deepseek-6.7B-Instruct-AWQ & 0.1864 & 1.609 & 1.762 & 0.2048 & 1.532 & 1.677 & 0.0972 & 1.472 & 1.608 \\
Deepseek-33B-Instruct-AWQ & 0.3149 & \textbf{1.719} & \textbf{1.952} & 0.3492 & \textbf{1.654} & \textbf{1.862} & 0.1834 & \textbf{1.576} & \textbf{1.776} \\
GPT-4o-mini-2024-07-18 & 0.5757 & 1.517 & 1.777 & 0.5870 & 1.358 & 1.548 & 0.3642 & 1.459 & 1.610 \\
GPT-4o-2024-08-06 & \textbf{0.6327} & 1.380 & 1.507 & \textbf{0.6263} & 1.351 & 1.482 & \textbf{0.4534} & 1.419 & 1.559 \\ \midrule
Human &  & 2.305 & 2.609 &  & 2.198 & 2.499 &  & 2.209 & 2.530 \\ \bottomrule
\end{tabular}%
}
\caption{Functional correctness and algorithmic diversity on the APPS dataset. The best results are marked in bold.}
\label{tab:div-analysis}
\end{table*}

\subsection{Diversity Analysis on Correct Solutions}
\begin{figure*}[t]
    \centering
    \begin{subfigure}[b]{0.32\textwidth}
        \centering
        \includegraphics[width=\textwidth]{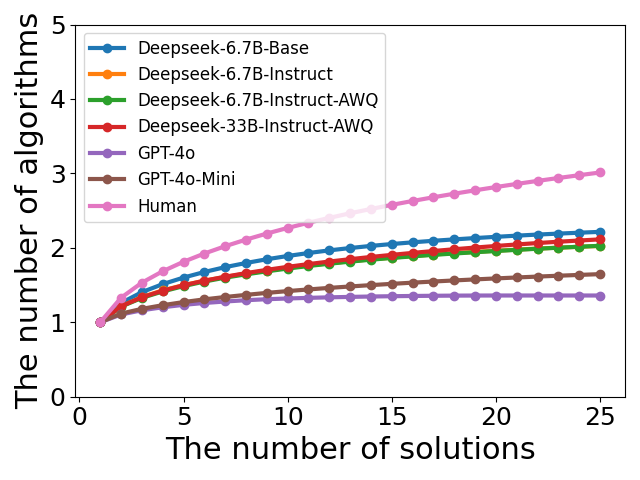}
        \caption{Introductory, AtCoder}
        \label{fig:adc_intro_atcoder}
    \end{subfigure}
    \hfill
    \begin{subfigure}[b]{0.32\textwidth}
        \centering
        \includegraphics[width=\textwidth]{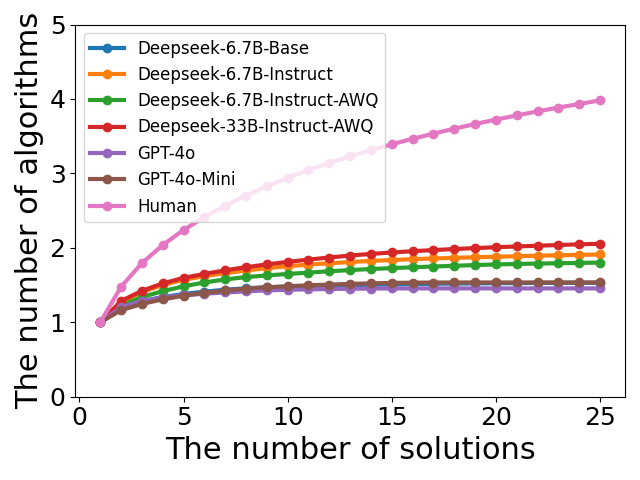}
        \caption{Interview, AtCoder}
    \end{subfigure}
    \hfill
    \begin{subfigure}[b]{0.32\textwidth}
        \centering
        \includegraphics[width=\textwidth]{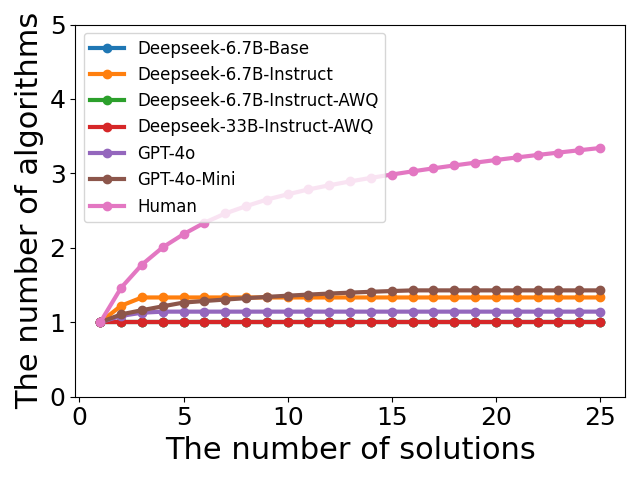}
        \caption{Competition, AtCoder}
    \end{subfigure}

    \begin{subfigure}[b]{0.32\textwidth}
        \centering
        \includegraphics[width=\textwidth]{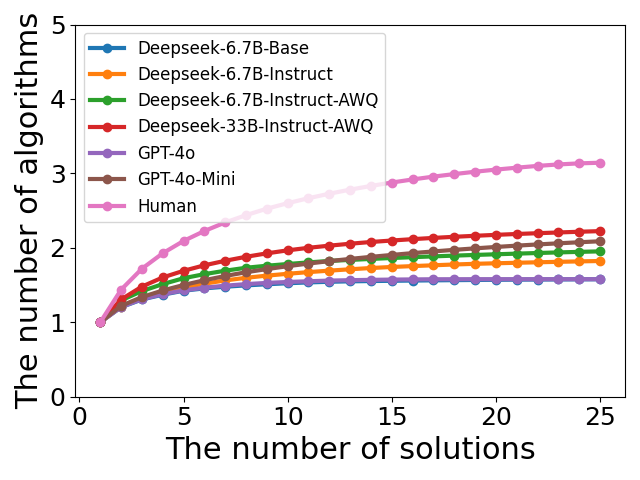}
        \caption{Introductory, CodeForces}
    \end{subfigure}
    \hfill
    \begin{subfigure}[b]{0.32\textwidth}
        \centering
        \includegraphics[width=\textwidth]{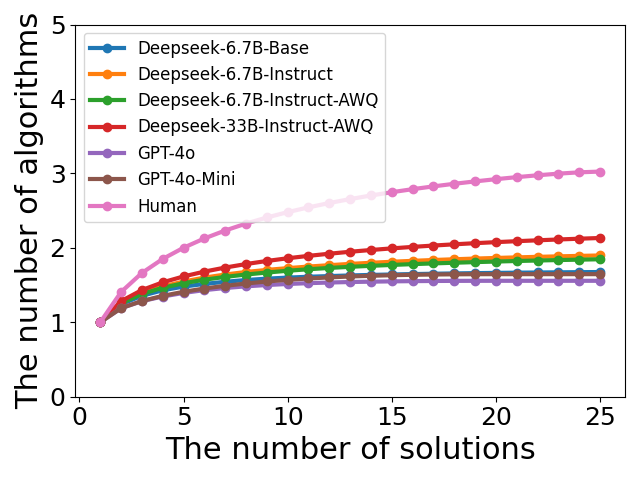}
        \caption{Interview, CodeForces}
    \end{subfigure}
    \hfill
    \begin{subfigure}[b]{0.32\textwidth}
        \centering
        \includegraphics[width=\textwidth]{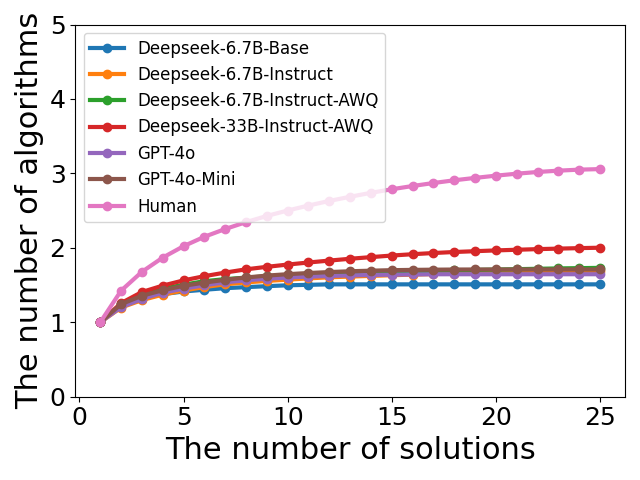}
        \caption{Competition, CodeForces}
    \end{subfigure}

    \caption{The Algorithmic Diversity Curve of the human solution set and model ones for the APPS dataset.}
    \label{fig:adc}
\end{figure*}

We analyze functionally correct solutions generated by models and collected solutions written by human from the perspective of algorithmic diversity.
Table~\ref{tab:div-analysis} reports the Pass@10 score with our proposed metrics, $\mathrm{EA}$ and $\mathrm{NAUADC}$, to examine the relationship between diversity and functional correctness across different solution sets.
In addition, we present the Algorithmic Diversity Curve in Figure~\ref{fig:adc}, highlighting key differences between human-written and model-generated solutions.

\paragraph{Findings 1. Model-generated solutions show low algorithmic diversity.}
Our algorithmic diversity curve clearly illustrates that model-generated solutions exhibit significantly less algorithmic variety than human-written solutions.
As shown in Figure~\ref{fig:adc_intro_atcoder}, humans can discover an average of three distinct algorithms, whereas the model develops only two for AtCoder-\textit{introductory} problems when generating 25 correct solutions on average ($\mathrm{DA@}25$).
This empirical findings are reflected in the $\mathrm{EA}$ and $\mathrm{NAUADC}$ scores in Table~\ref{tab:div-analysis}, where human-written solutions consistently achieve higher scores than model-generated ones.
This implies that even when we train other models with these model-generated solutions, they are unlikely to learn generalized problem-solving strategies due to the lack of algorithmic diversity encoded in these solution set.

\paragraph{Findings 2. Algorithmic diversity varies depending on problem semantics.}
We demonstrate that the problems from different sources and difficulty show different trends in terms of algorithmic diversity.
For AtCoder-\textit{introductory} problems, we observe that while using larger models like Deepseek-33B-Instruct-AWQ improves Pass@10 score, it did not lead to greater algorithmic diversity (Table~\ref{tab:div-analysis}).
However, there exists notable increment in algorithmic diversity scores for CodeForces problems.
Our analysis reveals that the AtCoder-\textit{introductory} problems typically require straightforward problem-solving approaches, while others demand more code-level algorithmic variability, particularly elicited from larger models.

\paragraph{Findings 3. Human solutions also contain redundant algorithms.}
In Table~\ref{tab:div-analysis}, human-written solutions mostly consist of two distinct algorithms per problem, despite humans' superior ability to explore diverse algorithmic approaches.
There are two possible explanations for this limited algorithmic diversity.
First, the problems may be too semantically straightforward to solve, suggesting there is little value in approaching them differently.
Second, the problems may be so challenging that humans have yet to discover alternative algorithms.
We speculate that the first explanation accounts for the lower diversity observed in the \textit{introductory} problems.
Through our manual inspection, we confirm that most human solutions in the \textit{introductory} problems converge on the same algorithm.
We attribute this phenomenon to the dataset originating from educational platforms designed to help students reinforce their programming skills by reimplementing well-known approaches.

\paragraph{Findings 4. GPT-4o series exhibits low algorithmic diversity.}
Our experimental results show that powerful proprietary models has low algorithmic diversity compared to their open-source counterparts regardless of problem complexity or the original source.
It implies that GPT-4o models are effective at finding one correct solution, but they struggle to identify alternative valid approaches.
In contrast, Deepseek models show relatively high diversity even though they produce less correct solution compared to GPT-4o models.
With these diverse solutions, we can derive the high-level abstract concept of how to solve a problem that cannot be acquired from one single solution.

\paragraph{Findings 5. The effect of instruction tuning on algorithmic diversity varies depending on problem difficulty.}
When comparing the Deepseek-base and Deepseek-instruct models, algorithmic diversity shows different patterns: instruction tuning decreases algorithmic diversity in AtCoder-\textit{introductory} problems, but increases it in other problems.
A key distinction between AtCoder-\textit{introductory} problems and others lies in their correctness scores.
Based on Pass@10 scores, most AtCoder \textit{introductory} problems are easily solvable by the models, whereas others pose significant challenges.
This suggests that instruction tuning favors single, straightforward answers for simple problems while promoting the exploration of multiple approaches for complex problems.

\begin{table}[t]
\centering
\resizebox{\columnwidth}{!}{%
\begin{tabular}{@{}ccccc@{}}
\toprule
Qwen & Deepseek & GPT & EA & NAUADC \\ \midrule
\multicolumn{5}{c}{Source: AtCoder} \\ \midrule
 & \checkmark & \checkmark & $1.597 \left(\textcolor{teal}{+0.049}\right)$ & $1.837 \left(\textcolor{teal}{+0.057}\right)$ \\
\checkmark &  & \checkmark & $1.357 \left(\textcolor{teal}{+0.013}\right)$ & $1.533 \left(\textcolor{teal}{+0.015}\right)$ \\
\checkmark & \checkmark &  & $1.550 \left(\textcolor{teal}{+0.002}\right)$ & $1.759 \left(\textcolor{purple}{-0.021}\right)$ \\
\checkmark & \checkmark & \checkmark & $1.563 \left(\textcolor{teal}{+0.015}\right)$ & $1.771 \left(\textcolor{purple}{-0.009}\right)$ \\ \midrule
\multicolumn{5}{c}{Source: CodeForces} \\ \midrule
 & \checkmark & \checkmark & $1.759 \left(\textcolor{teal}{+0.040}\right)$ & $1.980 \left(\textcolor{teal}{+0.028}\right)$ \\
\checkmark &  & \checkmark & $1.814 \left(\textcolor{teal}{+0.126}\right)$ & $2.093 \left(\textcolor{teal}{+0.168}\right)$ \\
\checkmark & \checkmark &  & $1.862 \left(\textcolor{teal}{+0.143}\right)$ & $2.141 \left(\textcolor{teal}{+0.189}\right)$ \\
\checkmark & \checkmark & \checkmark & $1.922 \left(\textcolor{teal}{+0.203}\right)$ & $2.196 \left(\textcolor{teal}{+0.244}\right)$ \\ \bottomrule
\end{tabular}%
}
\caption{Algorithmic diversity of a merged solution set on APPS \textit{introductory} problems. Qwen, Deepseek, and GPT stand for Qwen2.5-Coder-32B-Instruct-AWQ, Deepseek-coder-33B-Instruct-AWQ, and gpt-4o-2024-08-06, respectively. The numbers in parentheses indicate the improvement relative to the highest performance achieved by any individual model.}
\label{tab:solution_merge}
\end{table}

\subsection{Diversity Analysis on Merged Solutions}
One simple approach to increasing the diversity of solutions could be integrating solution sets derived from different models.
To validate the effectiveness of this approach, we merge solutions from different models and evaluate them in Table~\ref{tab:solution_merge}.

Our empirical investigations indicate that amalgamating solutions derived from diverse models yields varying degrees of diversity enhancement.
Primarily, the AtCoder problems show negligible enhancement or fail to augment diversity even upon the consolidation of their respective solutions.
It suggests that merging these solutions is not an effective approach to address the limited diversity.
Conversely, for CodeForces problems, we observe a substantial increase in diversity given a merged solution set.
This suggests that these models employ different approaches when implementing CodeForces challenges, enhancing the heterogeneity of the merged solution set.

Moreover, it is noteworthy that the choice of models incorporated into the merged set significantly impacts algorithmic diversity.
The integration of Deepseek and GPT solutions results in a relatively small increase in diversity compared to integrations involving Qwen solutions.
This can be attributed to the substantial similarity between Deepseek and GPT in their algorithmic foundations and development paradigms.
While merging all available models achieves the highest solution diversity, these findings emphasize the critical importance of selecting models with distinctive solution-generating capabilities, particularly when working with limited model combinations.

\begin{figure}[t]
    \centering
    \includegraphics[width=\linewidth]{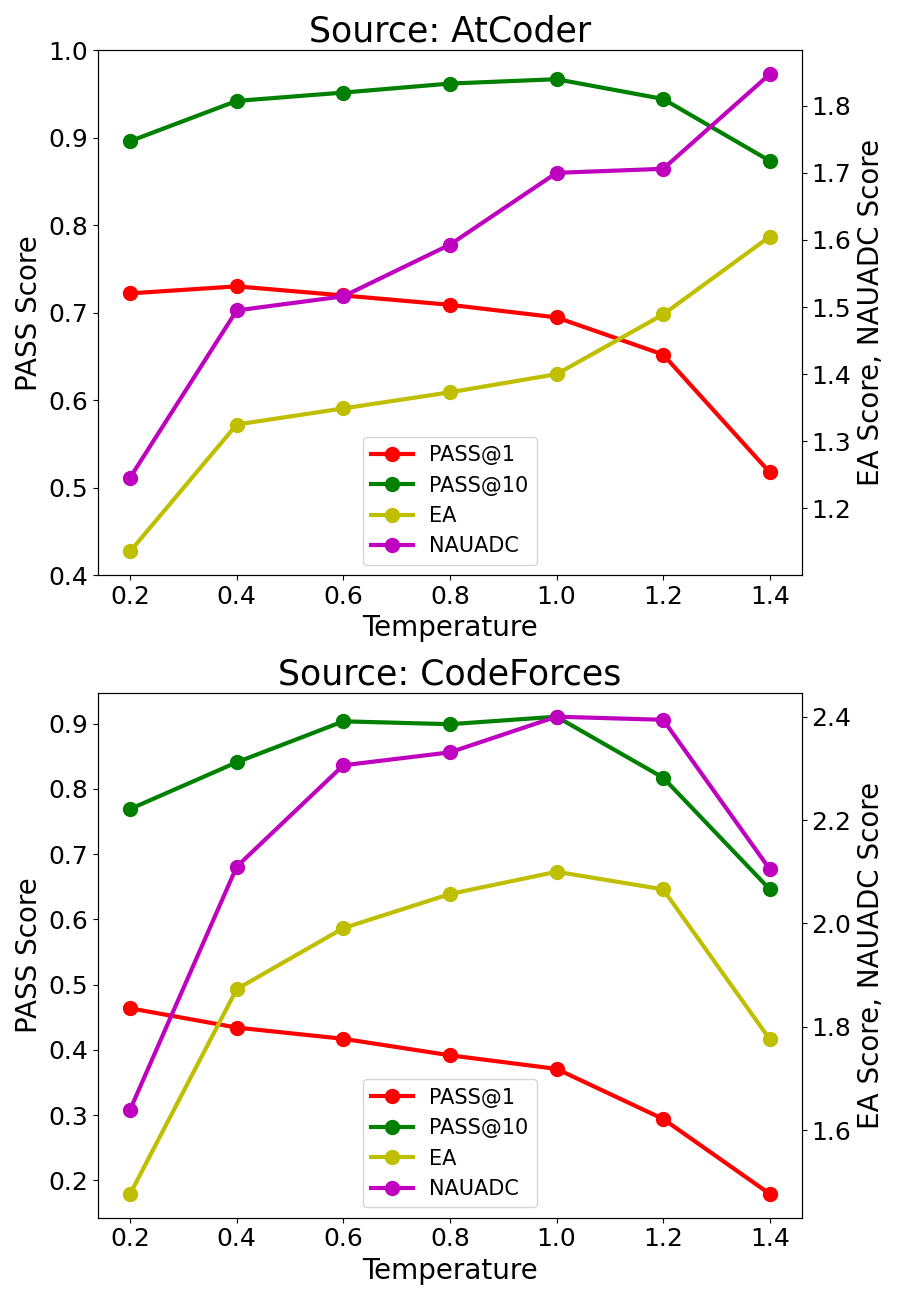}
    \caption{The impact of sampling temperature on correctness and algorithmic diversity. Dataset: APPS \textit{introductory}, Model: Deepseek-coder-33B-Instruct-AWQ.}
    \label{fig:temperature}
\end{figure}

\begin{table*}[t]
\small
\centering
\begin{tabular}{@{}ccc@{}}
\toprule
$\mathrm{AlgoSim}$ & Human & Completion \\ \midrule
Positive & Positive & \begin{tabular}[c]{p{0.75\linewidth}}To classify ... (omitted) ... \textcolor{teal}{The previous solution calculates the number of possible passwords by cubing the input number N, which is equivalent to N\textasciicircum{}3.} ... (omitted) ... \textcolor{purple}{The given solution also calculates the number of possible passwords by cubing the input number N, but it explicitly multiplies N by itself three times (N * N * N).} ... (omitted) ... Based on this analysis, the given solution leverages similar logic to the previous solution, but with a more explicit and step-by-step approach. Decision: similar to the previous solution.\end{tabular} \\ \midrule
Negative & Negative & \begin{tabular}[c]{p{0.75\linewidth}}To classify ... (omitted) ... \textcolor{teal}{The previous solution uses the formula N\textasciicircum{}3 to calculate the number of possible passwords. This is a mathematical approach that directly calculates the result without iterating over all possible combinations.} \textcolor{purple}{The given solution, on the other hand, uses a brute-force approach by iterating over all possible combinations of digits from 1 to N and counting them.} ... (omitted) ... Therefore, the given solution does not leverage similar logic to the previous solution. It uses a novel approach to solve the problem. Decision: a novel approach.\end{tabular} \\ \midrule
Positive & Negative & \begin{tabular}[c]{p{0.75\linewidth}}To classify ... (omitted) ... \textcolor{teal}{The previous solution calculates the number of possible passwords by cubing the input number N} ... (omitted) ... \textcolor{purple}{The given solution uses the itertools.product function to generate all possible combinations of 3 digits from 1 to N, and then prints the length of this list.} ... (omitted) ... Comparing the two solutions, we can see that the given solution uses a more explicit and iterative approach to generate all possible combinations, whereas the previous solution uses a mathematical formula to calculate the result directly. \textcolor{red}{However, both solutions are based on the same underlying logic, which is to count the number of possible combinations of 3 digits from 1 to N.} Therefore, the given solution leverages similar logic to the previous solution, but with a more explicit and iterative approach.
Decision: similar to the previous solution.\end{tabular} \\ \bottomrule
\end{tabular}
\caption{Qualitative examples from $\mathrm{AlgoSim}$ completion. We color the explanation for the first solution to \textcolor{teal}{teal} while those for the second solution to \textcolor{purple}{purple}. We point out the misleading sentence as \textcolor{red}{red}. The model should compare how to count the number of possible combinations.}
\label{tab:qualitative}
\end{table*}

\subsection{Sampling Temperature Analysis}
We investigate how the sampling temperature affects the algorithmic diversity of generated code.
We sample 50 problems from the APPS-\textit{introductory} problems, and generate 100 solutions per problem with the sampling temperatures ranging from 0.2 to 1.4 in increments of 0.2.
The functional correctness and diversity scores across different temperature settings are reported in Figure~\ref{fig:temperature}.
The results clearly illustrate the correlation between the sampling temperature and algorithmic diversity.
As the sampling temperature increases, both EA and $\mathrm{NAUADC}$ scores gradually rise.
However, when evaluating CodeForces problems at a temperature of 1.4, we observe a significant drop in $\mathrm{EA}$ and $\mathrm{NAUADC}$ scores, likely due to the model's reduced stability in generating correct solutions.

We emphasize that simply comparing Pass@1 and Pass@10 scores appears insufficient for understanding the diversity of model-generated solutions.
As the sampling temperature increases from 0.2 to 1.0, we observe an increase in Pass@10 scores accompanied by a decrease in Pass@1 scores, aligning with previous research.
Within this range, increasing the diversity of model generation based on high temperature increases the probability of finding the right solution;
that is, diversity positively correlates with accuracy.
Once the temperature exceeds 1.0, however, the Pass@10 begins to decline due to a higher prevalence of incorrect solutions.
Nevertheless, the algorithmic diversity of valid solutions continues to increase at the highest sampling temperatures.
Our study emphasizes the significance of high-temperature sampling, which has been largely overlooked due to existing evaluation methods solely focusing on accuracy.

\subsection{Qualitative analysis on $\mathrm{AlgoSim}$}
We analyze $\mathrm{AlgoSim}$ behavior of through our manual inspection.
Some representative examples are listed up in Table~\ref{tab:qualitative}.
We confirm that the logical progression of $\mathrm{AlgoSim}$ follows a four-step structure: (1) problem overview, (2) solution methodology analysis, (3) comparative logical assessment, and (4) conclusive determination.
We discover that while the model can effectively explain the logical progression of solutions, it struggles to identify which aspect should be compared in this context, e.g., the underlying logic or the intended purposes of those solutions.
In our prompt, we request the model to compare and analyze the \textit{logic} between the given two solutions, as they share the same objective of problem-solving.
We anticipate this issue will be naturally resolved as LLMs' reasoning capabilities continue to advance.

\begin{table}[]
\small
\centering
\begin{tabular}{@{}cccc@{}}
\toprule
Edit distance & $x \le 100$ & $100 < x \le 500$ & $500 < x$ \\ \midrule
Claude-3.5 & 0.60 & 0.75 & 0.70 \\
GPT-4o & 0.65 & 0.70 & 0.60 \\
Human & 0.91 & 0.78 & 0.72 \\
$\kappa$ & - & 0.58 & 0.25 \\ \bottomrule
\end{tabular}
\caption{$\mathrm{AlgoSim}$ acceptance ratio results. We additionally report survey responses from recent LLMs. $\kappa$ indicates Cohen's Kappa across human annotators.}
\label{tab:survey}
\end{table}

Furthermore, we conducted a human evaluation study to assess how $\mathrm{AlgoSim}$ effectively distinguishes algorithmic differences between two lexically different implementations.
We categorize solution pairs according to their lexical difference, quantified through edit distance, into 3 groups and extract 20 representative cases from each group.
We gathered survey responses to evaluate whether the participants agree with the AlgoSim's reasoning about algorithmic analysis and their predictions.
Further details can be found in Appendix~\ref{sec:human_study}.

We report survey statistics in Table~\ref{tab:survey}.
The acceptance ratio exceeded 0.6 on average, demonstrating their capability of distinguishing algorithmic difference.
One notable trend is that human acceptance and the corresponding agreement decrease as problem complexity increases.
This phenomenon is attributed to the inherently subjective nature of the concept of logic.
For instance, someone considers BFS and DFS algorithms as the same graph traversal logic while others don't.

\section{Conclusion}
\label{sec:conclusion}
Throughout this work, we comprehensively analyze the code diversity of model-generated solutions.
We develop clustering-based metrics, namely $\mathrm{DA@}K$, $\mathrm{EA}$, and $\mathrm{NAUADC}$, leveraging LLMs' capabilities to effectively contrast code semantics across different codes.
These metrics yield meaningful indicators, i.e., the mean algorithmic count within solution sets, while accounting for the distribution of generated solutions.
Our extensive analysis unveils previously unexplored findings about code diversity.
First, we identify the algorithmic redundancy inside model-generated solutions.
Especially, our algorithmic diversity curve reveals that the model struggles to devise novel algorithms, despite generating numerous correct solutions.
Finally, we investigate multiple methods to improve algorithmic diversity through solution merging and adjusting sampling temperature, showing that combining solutions from heterogeneous models and generating with extremely high sampling temperature effectively amplifies algorithmic variety.

We hope that our findings facilitate further research aimed at enhancing code diversity to generate more diverse and correct codes.


\section*{Limitations}
\label{sec:limitation}
We identify a few limitations during our experiments and provide promising research directions for facilitating research about code diversity.

First, even though we provide a robust estimator for $\mathrm{DA@}K$, it does not reflect code diversity properly if it produces less than two solutions for a problem, where it often occurs in challenging problems such as APPS-\textit{competition} problems.
Therefore, analyzing diversity with a reduced number of correct solutions or without them becomes a promising research topic.

Second, our diversity analysis heavily rely on reasoning ability in large language models, necessitating substantial computational resources.
Therefore, efficiently contrasting two codes with small language models specialized on analyzing their algorithm becomes a promising research direction.

Finally, our human study revealed that the interpretation of logic varies among individuals.
Although the participants establish their own standards of logic and provide consistent survey response, the criteria among participants occasionally lack global concordance.

\section*{Ethical consideration}
\label{sec:ethical}
In our research, we have exclusively utilized open-source language models (LMs) and datasets. 
This ensures transparency and allows the broader research community to replicate and build upon our work without legal or proprietary constraints. 
Also, we recruited graduate students in computer science to evaluate code similarity, ensuring their involvement was voluntary and informed. 
We do not intend harm or overlook significant ethical issues.

\bibliography{anthology,ref}
\bibliographystyle{acl_natbib}

\appendix
\section{Potential Risks}
In this section, we identify potential risks and systematically elaborate on their precautions.
First, the model may produce security-vulnerable code that could be potentially harmful to the user's system.
This risk can be mitigated by executing model-generated code within a sandboxed environment.
Additionally, the analysis results from $\mathrm{AlgoSim}$ may lead to biased conclusions among researchers due to the inherent bias in language models~\citep{liu-etal-2023-g}.
We anticipate that this risk will naturally diminish as future language models evolve to become unbiased.

\section{Scientific Artifacts}
We leverage publicly available datasets with proper citation while maintaining legal compliance and ethcial use.
While the APPS dataset is licensed under CC BY-SA 3.0, the original research publication acknowledges potential copyright considerations for problems sourced from AtCoder and CodeForces platforms.
The authors cite Fair Use \S 107 to justify their utilization of these problems.
In accordance with these principles, our research-oriented use of this dataset adheres to the Fair Use principles, thereby ensuring legal compliance.
The researchers designated their test dataset specifically for evaluation purposes, and we complied with their intended usage by analyzing various language models for assessment rather than training.
Furthermore, the creators ensured that their dataset excluded any personally identifiable information or inappropriate content.

\section{Computational Experiments}
In this research, we employed a combination of open-source and proprietary models, with parameter sizes ranging from 6.7 to 33 billion.
While we cannot report the parameter counts of proprietary models because they do not disclose that information, we provide their release dates to facilitate the replication of our study.
The experimental computations consumed approximately 500 A100-80G GPU hours, primarily due to the extensive code solution sampling required per problem and the quadratic computational complexity involved in comparing algorithmic differences between code implementations.
In addition, we spend more than 1000 dollars to sample solutions from proprietary models.
We intend to make the completions publicly available to ensure accessibility for users who may face budget constraints in accessing these resources.

\section{Human Study}
\label{sec:human_study}
We conduct a human study to measure the quality of $\mathrm{AlgoSim}$.
In this section, we provide a detailed experimental setup to replicate our experiments.

\begin{table}[t]
\small
\centering
\begin{tabular}{p{0.9\columnwidth}}
\toprule
In the given file, there are solution1 and solution2 for the same problem in one row. \\\\
Rationale is a column that stores the LLM's reasoning results of whether the two solutions use the same logic or not to solve the given problem. \\\\
Your task is to look at the two codes and the rationale results, and if you agree with the model's idea, write "Agree", if you do not agree, write "Disagree" and briefly describe the reason in less than two lines. \\\\
All codes passed the test case. \\
\bottomrule
\end{tabular}
\caption{Instruction given to the participant.}
\label{tab:human_inst}
\end{table}

\subsection{Instruction for Human Study}
The primary objective of our human evaluation study is to assess the acceptability of $\mathrm{AlgoSim}$'s predictions based on human reasoning.
To facilitate this assessment, we present participants with a problem statement, two lexically different solutions, and $\mathrm{AlgoSim}$'s output which encompasses both reasoning and predictions.
We instructed evaluators to verify the credibility of the output, ensuring it contains no fabricated or erroneous information that could lead to incorrect conclusions.
We present our instruction in human study in Table~\ref{tab:human_inst}.

\subsection{Information about Human Participants}
We recruited eight participants (four undergraduate and four graduate students) who possessed intermediate programming skills and demonstrated sufficient understanding the solution logic of coding problems used in our study.
Prior to commencing our human study, we obtain consent from participants regarding response collection.
If participants decline consent, their responses will not be recorded.
Participants who consent to data collection will receive a Starbucks gift card valued at \$8 as compensation, which constitutes legitimate remuneration in the country where the study was conducted.

\section{AI Assistants in Research}
We actively utilize Copilot through Visual Studio Code editor for implementing our code for the experiments.
All codes have been verified by the authors and thoroughly tested.
Furthermore, we initially authored this paper and refined our writing with the assistance of Claude, an AI language model, to effectively communicate our research findings to the academic community.

\begin{table}[t]
\small
\centering
\begin{tabular}{p{0.9\columnwidth}}
\toprule
Your task is to classify whether a given solution solves a problem with similar logic to existing solutions or whether it leverages a novel approach.\\\\
You will be given a problem and a previous solution that has been used to solve that problem. If the given solution leverages similar logic to the previous solution, conclude your response with the sentence "Decision: similar to the previous solution." Otherwise, conclude your response with the sentence "Decision: a novel approach." Include your reasoning for performing this task in your response. \\\\
Below, the problem is provided wrapped in the <|PROBLEM|> tag, the previous solution is provided wrapped in the <|PREVIOUS SOLUTION|> tag, and the solution to be classified is provided within the <|SOLUTION|> tag.\\\\

<|PROBLEM|>\\
\{question\}\\
<|/PROBLEM|>\\\\

<|PREVIOUS SOLUTION|>\\
\{past\_solution\}\\
<|/PREVIOUS SOLUTION|>\\\\

<|SOLUTION|>\\
\{solution\}\\
<|/SOLUTION|>\\
\bottomrule
\end{tabular}
\caption{Query for inferring algorithmic difference using instruction-tuned LLMs.}
\label{tab:prompt}
\end{table}

\section{Prompt Examples}
\label{app:prompt}
Our query for evaluating the algorithmic similarity is presented in Table~\ref{tab:prompt}.

\section{$\mathrm{AlgoSim}$ Examples}
To facilitate comprehension of $\mathrm{AlgoSim}$, we present several illustrative examples in this section.
The complete set of examples will be available in our Hugging Face repository.

\begin{table*}[]
\small
\centering
\begin{tabular}{@{}l@{}}
\toprule
\textbf{Problem statements} \\ \midrule
\begin{tabular}[c]{@{}l@{}}You are given a three-digit positive integer N.\\ \\ Determine whether N is a palindromic number.\\ \\ Here, a palindromic number is an integer that reads the same backward as forward in decimal notation.\\ \\ -----Constraints-----\\  - $100 \le N \le 999$\\  - N is an integer.\\ \\ -----Input-----\\ Input is given from Standard Input in the following format:  \\ N\\ \\ -----Output-----\\ If N is a palindromic number, print Yes; otherwise, print No.\\ \\ -----Sample Input-----\\ 575\\ \\ -----Sample Output-----\\ Yes\\ N=575 is also 575 when read backward, so it is a palindromic number. You should print Yes.\end{tabular} \\ \midrule
\textbf{Solution1} \\ \midrule
\begin{lstlisting}
N = input()
if N == N{[}::-1{]}:
  print("Yes")
else:
  print("No")
  
\end{lstlisting} \\ \midrule
\textbf{Solution2} \\ \midrule
\begin{lstlisting}
N = int(input())
print('Yes' if str(N) == str(N){[}::-1{]} else 'No')

\end{lstlisting} \\ \midrule
\textbf{Completion} \\ \midrule
\begin{tabular}[c]{@{}p{\textwidth}@{}}To classify whether the given solution solves the problem with similar logic to the existing solution or leverages a novel approach, I will analyze the provided problem, previous solution, and the solution to be classified.\\ \\ The problem requires determining whether a given three-digit positive integer N is a palindromic number. The previous solution uses Python and checks if the input string N is equal to its reverse using slicing (`N{[}::-1{]}`). If they are equal, it prints "Yes"; otherwise, it prints "No".\\ \\ The solution to be classified also uses Python and checks if the input integer N is equal to its reverse by converting it to a string and then comparing it with its reverse. If they are equal, it prints "Yes"; otherwise, it prints "No".\\ \\ Upon analyzing both solutions, I notice that they share a similar approach to solving the problem. Both solutions use string manipulation to reverse the input and compare it with the original input. The main difference lies in the way they convert the integer to a string, with the previous solution using slicing (`N{[}::-1{]}`) and the solution to be classified using the `str()` function.\\ \\ However, the underlying logic of comparing the input with its reverse remains the same in both solutions. Therefore, I conclude that the solution to be classified leverages similar logic to the previous solution.\\ Decision: similar to the previous solution.\end{tabular} \\ \bottomrule
\end{tabular}
\caption{$\mathrm{AlgoSim}$ example (easy). Character-level edit distance between two solutions: 52}
\end{table*}

\begin{table*}[]
\small
\centering
\begin{tabular}{@{}l@{}}
\toprule
\textbf{Problem statements} \\ \midrule
\begin{tabular}[c]{@{}l@{}}It is September 9 in Japan now.\\ You are given a two-digit integer N. Answer the question: Is 9 contained in the decimal notation of N?\\ \\ -----Constraints-----\\  - $10 \le N \le 99$\\ \\ -----Input-----\\ Input is given from Standard Input in the following format:\\ N\\ \\ -----Output-----\\ If 9 is contained in the decimal notation of N, print Yes; if not, print No.\\ \\ -----Sample Input-----\\ 29\\ \\ -----Sample Output-----\\ Yes\\ The one's digit of 29 is 9.\end{tabular} \\ \midrule
\textbf{Solution1} \\ \midrule
\begin{lstlisting}
N = int(input().strip())
print('Yes' if '9' in str(N) else 'No')
\end{lstlisting} \\ \midrule
\textbf{Solution2} \\ \midrule
\begin{lstlisting}
N = int(input())
if N % 10 == 9 or N // 10 == 9:
    print('Yes')
else:
    print('No')
\end{lstlisting} \\ \midrule
\textbf{Completion} \\ \midrule
\begin{tabular}[c]{@{}p{\textwidth}@{}}To classify whether the given solution leverages similar logic to the previous solution or a novel approach, I will analyze the problem and the solutions provided.\\ \\ The problem is to determine if the digit 9 is contained in the decimal notation of a two-digit integer N. The constraints are that N is a two-digit integer between 10 and 99, inclusive.\\ \\ The previous solution uses a straightforward approach by converting the integer N to a string and checking if the character '9' is present in the string. This approach is simple and effective for this problem.\\ \\ The given solution, on the other hand, uses a different approach. Instead of converting the integer to a string, it checks if the last digit (obtained by N \% 10) or the first digit (obtained by N // 10) is equal to 9. This approach is based on the fact that a two-digit integer can be represented as 10a + b, where a is the first digit and b is the last digit. Therefore, checking if a or b is equal to 9 is equivalent to checking if 9 is contained in the decimal notation of N.\\ \\ Based on this analysis, the given solution leverages a novel approach compared to the previous solution. The previous solution uses a string-based approach, while the given solution uses a mathematical approach based on the representation of a two-digit integer.\\ \\ Decision: a novel approach.\end{tabular} \\ \bottomrule
\end{tabular}
\caption{$\mathrm{AlgoSim}$ example (easy). Character-level edit distance between two solutions: 55}
\end{table*}

\begin{table*}[]
\small
\centering
\begin{tabular}{@{}l@{}}
\toprule
\textbf{Solution1} \\ \midrule
\begin{lstlisting}[basicstyle=\ttfamily\tiny]
from collections import deque
 
def solve(n):
  q = deque([(n, 0)]) # queue to store the current value and its corresponding moves
  visited = set() # set to store the visited values
  
  while q:
    cur, step = q.popleft()
  
  if cur == 1: # If the current value is 1, then return the moves
    return step
  
  if cur in visited: # If the current value has been visited, then skip it
    continue
    visited.add(cur)
  
  if cur % 6 == 0: # If the current value is divisible by 6, then add two new states to the queue
    q.append((cur // 6, step + 1))
    q.append((cur * 2, step + 1))
  elif cur % 3 == 0: # If the current value is divisible by 3, then only add one new state to the queue
    q.append((cur * 2, step + 1))
  else: # If the current value is not divisible by 3, then it is not possible to get 1
    return -1
 
# Read the number of test cases
t = int(input().strip())
 
# Solve each test case
for _ in range(t):
  n = int(input().strip())
  print(solve(n))
\end{lstlisting} \\ \midrule
\textbf{Solution2} \\ \midrule
\begin{lstlisting}[basicstyle=\ttfamily\tiny]
from collections import deque
 
# Check if a number is valid
def is_valid(n):
  return 1 <= n <= 10**9
 
# Perform operations
def perform_operations(n):
  multiply_by_two = n * 2
  divide_by_six = n // 6 if n % 6 == 0 else -1
  return [multiply_by_two, divide_by_six]
 
# BFS algorithm
def bfs(n):
  visited = {n}
  queue = deque([(n, 0)])
  while queue:
    node, level = queue.popleft()
  if node == 1:
    return level
  for next_node in perform_operations(node):
    if is_valid(next_node) and next_node not in visited:
      visited.add(next_node)
      queue.append((next_node, level + 1))
  return -1
 
# Read the number of test cases
t = int(input())
 
for _ in range(t):
  # Read the integer n
  n = int(input())
  # Perform BFS from n to 1
  print(bfs(n))
\end{lstlisting} \\ \midrule
\textbf{Completion} \\ \midrule
\begin{tabular}[c]{@{}p{\textwidth}@{}}The given problem is to find the minimum number of moves needed to obtain 1 from a given integer $n$ by either multiplying $n$ by two or dividing $n$ by 6 (if it is divisible by 6 without the remainder). The task is to determine if it's possible to do that or not. \\
The previous solution uses a breadth-first search (BFS) algorithm with a queue to store the current value and its corresponding moves. It also uses a set to store the visited values to avoid revisiting them. \\
The given solution also uses a BFS algorithm with a queue to store the current value and its corresponding moves. However, it uses a different approach to perform the operations. Instead of directly multiplying or dividing the current value, it uses a function \`perform\_operations\` to generate the next possible values. This function returns a list of two values: the result of multiplying the current value by 2, and the result of dividing the current value by 6 if it is divisible by 6 without the remainder, or -1 otherwise. \\
The given solution also uses a function \`is\_valid\` to check if a number is within the valid range. \\
Upon analyzing the given solution, it appears to leverage a similar approach to the previous solution, but with a different implementation of the operations. The logic of using a BFS algorithm with a queue and a set to store the visited values is similar, but the way the operations are performed is different. \\
Therefore, the decision is: \\
Decision: similar to the previous solution.\end{tabular} \\ \bottomrule
\end{tabular}
\caption{$\mathrm{AlgoSim}$ example (hard). Character-level edit distance between two solutions: 743}
\end{table*}

\end{document}